\documentclass{article}

\usepackage[letterpaper,margin=1.0in]{geometry}  

\usepackage{graphicx, amsmath,color,amsfonts}
\usepackage{authblk}
\usepackage[inline]{enumitem}
 \setitemize{noitemsep} 

\begin{document}

\title{Equation level matching: An extension of the method
  of matched asymptotic expansion for problems of wave
  propagation}

\author[1]{Luiz M. Faria}
\author[1]{Rodolfo R. Rosales}
\affil[1]{Department Mathematics,
  Massachusetts Institute of Technology,
  Cambridge, MA, USA, 02139}

\maketitle

\begin{abstract}
 We introduce an alternative to the method of matched asymptotic
 expansions. In the ``traditional'' implementation, approximate
 solutions, valid in different (but overlapping) regions are matched
 by using ``intermediate'' variables. Here we propose to match at the
 level of the equations involved, via a ``uniform expansion'' whose
 equations enfold those of the approximations to be matched.
 This has the advantage that one does not need to explicitly solve the
 asymptotic equations to do the matching, which can be quite impossible
 for some problems. In addition, it allows matching to proceed in
 certain wave situations where the traditional approach fails because
 the time behaviors differ (e.g., one of the expansions does
 not include dissipation).
 On the other hand, this approach does not provide the fairly explicit
 approximations resulting from standard matching. In fact, this is not
 even its aim, which to produce the ``simplest'' set of equations that
 capture the behavior.
\end{abstract}

\section{Introduction}
The roots of the method of matched asymptotic expansion can be traced
back to the seminal work by Prandtl on viscous boundary
layers~\cite{prandtl1904uber}. With a revolutionizing idea, Prandtl
realized that the effects of viscosity on an object moving through a
fluid are felt only in a thin layer near the object, where the fluid
velocity must rapidly match that of the object's surface. In Prandtl's
physical picture the flow encompasses two distinct domains: an outer
region, away from the object, where viscosity is negligible, and a thin
transition layer, where the fluid quickly changes its velocity to match
that of the moving object. A few years later G. I. Taylor elucidated
some puzzles surrounding shock waves by employing a  modification of
Prandtl's boundary layer concept --- an internal, moving,
thin layer~\cite{Taylor1910dmg} where the fluid rapidly change values.
%

The method of 
of matched asymptotic expansions was given a solid mathematical foundation
in the 1950's, through the work of Kaplun and Lagerstrom~\cite{Kaplun1954,%
Kaplun1957,KaplunsPapers1967,KaplunLagerstrom1957}. Kaplun introduced a
precise definition for the 
``matching'' procedure for asymptotic approximations with overlapping
domain of validity. For a review of his ideas
see~\cite{LagerstromCasten1972}, as well as any of the many textbooks in
asymptotic expansions that describe it, such
as~\cite{BenderOrszag1978,ColeBook1968,MHHolmes1995,JKHunter2004,%
NayfehPerTech1981,VanDyke1975,kevorkian1996multiple}.
A rough, quick description, of the method follows. As an example, assume
that the objective is to obtain the solution to a boundary value problem
for an ODE, or a PDE, which has a small parameter in it. Then, first one
obtains several expansions for the solution of the problem, under various
scalings for the independent variables. Each expansion is valid in some
region of space, and may have parameters that need to be determined (e.g.,
to satisfy the boundary conditions). If the solutions have overlapping
regions of validity as the small parameter vanishes, then they can be
``matched'' by re-expanding each of them in terms of some intermediate
scaling, and equating the result term by term. This provides relationships
between the parameters that allow their determination, thus arriving at a
complete description of the solution. A final step, not always taken,
is to combine the various expressions for the solution (each valid in
some region) to produce a ``composite'' or ``uniform'' solution, which
describes the solution everywhere. In some sense, \emph{the composite
solution is the ``simplest'' approximation that captures the whole
behavior of the solution.} The aim of the method introduced in this paper
can be said to be: \emph{provide the ``simplest'' equation, or set, that
capture the whole behavior of the solutions.} The two concepts are
related, but they are not equal, as we will see though examples.

The method of matched asymptotic expansions is very powerful, and has
been successfully used to solve many problems in applications. For
example:
the theory of high activation energy asymptotics for flames
\cite{BuckmasterLudford1982},
viscous flows past solid objects~\cite{LagerstromCole1955,%
SchlichtingKlaus2017,VanDyke1975,VanDyke1962,Zeytounian2004},
critical layers in parallel shear flows~\cite{benney1969new},
transonic flows~\cite{CookP1993},
freezing/melting interfaces and (more generally) heat transfer
problems~\cite{AzizNa1984,Cess1966},
physics of plasmas~\cite{BenilovThomas2014},
%
%
electro-chemistry and
electro-osmosis~\cite{ChuBazant2005,SquiresBazant2004}, etc.
However, it has limitations, particularly for problems that involve wave
motion. The matching process requires detailed information about the
solution, which may not be accessible. It is also, mainly, focused on
``one'' solution, not whole classes of possible time evolutions resulting
from varying initial conditions --- though it is very well suited for the
calculation of eigenmodes. For problems where the time evolution is
affected by, say, dissipation that occurs mainly on a boundary layer [but
not everywhere], the time evolution that the various expansions yield is
different for different regions --- hence a ``matching'' as described
above does not seem possible (for an example, see
\S~\ref{sec:example-from-acoust}, in particular \S~\ref{LAIC:sub:ELM1}).

To ameliorate the difficulties outlined above, we propose an extension of
the method of matched asymptotic expansions, which relies on matching by
using equations, not solutions: i.e., \emph{equation level matching.} To
explain the idea, imagine that two [or more] different expansions
(corresponding to different scalings) have been produced, each valid in
some region. Each expansion is characterized by the sequence of equations
that determine the terms in the expansion. In the approach proposed here,
instead of solving these equations, another expansion is constructed
(the uniform equations expansion), which must have the following
\textbf{property}:
\begin{equation} \label{pr:property-1}
 \parbox{0.85\textwidth}{\emph{Upon using in this uniform equations
  expansion the scalings that correspond to each of the expansions to
  be matched, re-expanding and collecting equal orders, the expansions
  to be matched are recovered.}}
\end{equation}
The process is best illustrated with examples, starting with a very simple
one \S~\ref{sec:asympt-match-equat}. The following points will become
clear throughout this paper: \vspace*{-0.2em}
\begin{enumerate}
 \item 
 The ``output'' of this method are equations whose solutions provide
 uniform approximations to the problem solution. These uniform
 approximations are similar to the ``uniform (or composite)
 solutions''~\cite{BenderOrszag1978,JKHunter2004,NayfehPerTech1981}
 that the standard method produces [at least for problems where both
 approaches work], but not exactly the same (see the example in
 \S~\ref{sec:asympt-match-equat}).
 \item \vspace*{-0.2em} 
 The uniform equation expansion is not a ``standard'' expansion, in the
 sense that the small parameter appears in the equations that characterize
 each level of the approximation. This must be so for the expansion to
 actually be uniform. This makes the approach more complicated, but it
 gives it the  added flexibility needed to deal with time dependence in
 wave problems.
 \item \vspace*{-0.2em} 
 The method is not a replacement for the standard method. For one, in
 problems where both approaches work, the standard method usually is
 simpler. In addition, its objective is to show that a particular set of
 equations provides a uniform expansion. Hence, for wave problems at
 least, it is a way to obtain canonical equations that model the simplest
 setting where a particular set of phenomena matter. But it does not
 actually provide any solutions. Finally, while designed to deal with
 problems where the standard approach has difficulties, there may very
 well be situations where the opposite is true --- though we have not
 investigated this possibility, we suspect that it happens.
\end{enumerate}
In the following sections we first present the method through a simple
example \S~\ref{sec:asympt-match-equat}, and then show its application to
more complex problems of wave propagation (a weakly nonlinear detonation
model in \S~\ref{sec:SWND}, and boundary layer dissipation for acoustic
waves in \S~\ref{sec:example-from-acoust}). We make no attempt at
presenting a ``general formulation'' of the method for nonlinear
problems (as we do for linear problems), but a specific example is
discussed.
%


\section{Asymptotic matching of equations: ODE simple BVP}
\label{sec:asympt-match-equat}
%
%
%
We illustrate the method with an ordinary differential equation (ODE)
boundary value problem (BVP) whose solution has a boundary layer. That is
\begin{equation}\label{eq:simple-bvp}
 \left[ \epsilon\,\tfrac{d^2}{d\/x^2} - \tfrac{d}{d\/x} + 1 + \epsilon
 \right] u  = 0\/,
 \quad  0<x<1\/, \;\; 0 < \epsilon \ll 1\/,
\end{equation}
with boundary conditions (BC) $u(0) = 1\/$ and $u(1) = 0\/$. As
$\epsilon \downarrow 0\/$, $u\/$ develops a steep layer\,\footnote{\sf
   \,We skip many details here, as we assume that the readers are
   familiar with the standard theory of matching.}
near $x=1\/$. A very simple problem, meant as an illustrative example
only.
%
%
\subsection{Standard matching}\label{sec:standard-matching}
The \emph{regular} or \emph{outer expansion} (can only satisfy the left
BC) is \vspace{-0.2em}         
\begin{equation} \label{eq:outer-expansion-sequence-of-problems}
 u \sim \mbox{$\sum_{n=0}^{\infty}$}\,\epsilon^n\,\tilde{u}_n(x)\/,
 \quad\mbox{with}\quad
 \tilde{\mathcal L} \tilde{u}_n = [d_x^2 + 1] \tilde{u}_{n-1}\/,
\end{equation}
where $d_x = \frac{d}{d\/x}$, $\tilde{\mathcal L} = d_x -1\/$, and
$\tilde{u}_n = 0\/$ for $n<0\/$. In terms of the ``inner'' variable
$y = (x-1)/\epsilon$, the equation becomes \vspace*{-0.2em}
\begin{equation}
 \left[ \tfrac{1}{\epsilon} \tfrac{d^2}{d\/y^2} -
 \tfrac{1}{\epsilon} \tfrac{d}{d\/y} + 1 + \epsilon \right] u = 0\/.
\end{equation}
The \emph{inner} or \emph{singular expansion} is then \vspace*{-0.2em}
\begin{equation} \label{eq:inner-expansion-sequence-of-problems}
 u \sim \mbox{$\sum_{n=0}^\infty$}\,\epsilon^n\,\hat{u}_n(y)\/,
 \quad\mbox{with}\quad
 \hat{\mathcal L} \hat{u}_n = - \hat{u}_{n-1} - \hat{u}_{n-2}\/,
\end{equation}
where $d_y = \frac{d}{d\/y}$, $\hat{\mathcal L} = d_y^2 - d_y\/$, and
$\hat{u}_n = 0\/$ for $n<0\/$.

The expansion in \eqref{eq:outer-expansion-sequence-of-problems} is valid
everywhere, except for too close to $x = 1\/$ (i.e.: as long as
$d_x\,u = \mathcal{O}(1)\/$). The expansion in
\eqref{eq:inner-expansion-sequence-of-problems} is valid as long as $y\/$
is not too large. Both expansions have an overlap region, which can be
captured by the variable $z = (x-1)/\delta\/$, where
$\epsilon \ll \delta \ll 1\/$ --- e.g., $\delta = \sqrt{\epsilon}\/$.
The standard matched asymptotic expansions procedure is to write both
expansions in terms of $z\/$, re-expand in terms of the new small
parameters, and require that they agree order by order --- thus figuring
out the values of any free constants that may appear in the expansions.
Here \eqref{eq:outer-expansion-sequence-of-problems}
yields \vspace*{-0.2em}
\begin{equation} \label{eq:solution-outer-expansion-simple-ode}
 \tilde{u}_0 = e^x\/, \quad \tilde{u}_1 = 2\,x\,e^x\/, \quad
 \tilde{u}_2 = (2\,x^2 + 4\,x)\,e^x\/, \quad \dots
\end{equation}
where $u(0)=1\/$ has been enforced. Similarly, upon enforcing $u(1)=0\/$,
\eqref{eq:inner-expansion-sequence-of-problems} yields \vspace*{-0.2em}
\begin{equation} \label{eq:solution-inner-expansion-simple-ode}
 \hat{u}_0 = a_0\,v_0\/, \;\; \hat{u}_1 = a_0\,v_1 + a_1\,v_0\/, \;\;
 \hat{u}_2 = a_0\,v_2 + (a_0 + a_1)\,v_1 + a_2\,v_0\/, \; \dots
\end{equation}
where the $a_n\/$ are constants and \vspace*{-0.2em}
\begin{equation} \label{eq:solution-inner-expansion-simple-ode-aux}
 v_0 = 1- e^y\/, \quad v_1 = y + y\,e^y\/, \quad v_2 =
 y + \tfrac{1}{2}\,y^2 + (y-\tfrac{1}{2}\,y^2)\,e^y\/, \quad \dots
\end{equation}
The matching procedure then yields $a_0 =e\/, a_1 =2\,e\/, a_2 =6\,e\/$,
etc.
%
\subsection{Equation matching}\label{SEIM::SBVP:sss:EqMa}       
The matching procedure in \S~\ref{sec:standard-matching} relies on being
able to solve the expansion's equations --- or, at least, have detailed
information about their solutions. What if this is not possible, as can
be the case for complex, time-dependent problems (e.g., wave
propagation)? Let us pretend this is the case here. Then, instead of
solve and match, we propose to construct another expansion, including
both the inner and outer problems as limits. Hence we will operate at
the equation level only. The matching expansion corresponds to the
\emph{composite or uniform solution} in standard matching, hence we
call it the \emph{uniform expansion.}

In the simple case in this section the uniform expansion is\vspace*{-0.2em}
\begin{equation}\label{eq:uniform-expansion-sequence-of-problems}
 u \sim \mbox{$\sum_{n=0}^\infty$}\,\epsilon^n\,u_n(x;\epsilon)\/, \quad
 \mbox{with} \quad {\mathcal L}\,u_n = -u_{n -1},
\end{equation}
where ${\mathcal L} = \epsilon\,d_x^2 - d_x + 1\/$, $u_n = 0\/$ for
$n<0\/$, $u_0(0) = 1\/$, and $u_0(1) = u_n(0) = u_n(1) = 0\/$ for
$n \geq 1\/$. The operator ${\mathcal{L}}\/$ follows by looking at
the terms involving $\tilde u_n\/$ in
\eqref{eq:outer-expansion-sequence-of-problems}, those
involving $\hat u_n$ in
\eqref{eq:inner-expansion-sequence-of-problems}, and writing the operator
that involves all of them.
This process is the equation level analog of the standard matching
composite solution, which is obtained by adding the inner and outer
solutions, and subtracting the common terms. However, analog does
not mean equivalent --- see \S~\ref{sec:comp-betw-meth}

Next we check that property (\ref{pr:property-1}) is satisfied. First,
expand each $u_n$ in \eqref{eq:uniform-expansion-sequence-of-problems}
using the outer scaling $u_n\sim\sum_{j=0}^\infty\,\epsilon^j\,u_{n,j}(x)\/$.
This yields\vspace*{-0.2em}
\begin{align} \label{eqn:uniform-to-outer-expansion}
  d_x^2\,u_{n,j-1} - d_x\,u_{n,j} + u_{n,j} = -u_{n-1,j}\/,
\end{align}
where $u_{n,j}=0\/$ if any subscript is  negative. Define
$U_m = \sum_{n+j = m}\,u_{n,j}\/$. Then, from 
(\ref{eqn:uniform-to-outer-expansion}) it follows that $\{U_m \}\/$
satisfies \eqref{eq:outer-expansion-sequence-of-problems}. Next expand
each term in \eqref{eq:uniform-expansion-sequence-of-problems} using
the scaling for \eqref{eq:inner-expansion-sequence-of-problems}:
$u_n\sim\sum_{j=0}^\infty\,\epsilon^j\,u_{n,j}(y)\/$.
This yields\vspace*{-0.2em}
\begin{align} \label{eqn:uniform-to-inner-expansion}
  d_y^2\,u_{n,j} - d_y\,u_{n,j} + u_{n,j-1} = -u_{n-1,j-1}\/.
\end{align}
Then, if $U_m = \sum_{n+j=m}\,u_{n,j}\/$, $\{ U_m \}\/$ satisfies
\eqref{eq:inner-expansion-sequence-of-problems}. Thus
\eqref{eq:uniform-expansion-sequence-of-problems} contains both the
inner and outer expansions.

For this example the leading order,
${\mathcal L}\,u_0 = (\epsilon\,d_x^2 - d_x + 1)\,u_0 = 0\/$  is not
substantially simpler than the original problem \eqref{eq:simple-bvp}.
This is not surprising when starting from a simple equation; for then
there is nothing fundamentally simpler that can approximate the full
behavior. One may then ask the question: \emph{is there any advantage for
the approach in \S~\ref{SEIM::SBVP:sss:EqMa} versus the one in
\S~\ref{sec:standard-matching}?} If the aim is to obtain explicit
approximations, certainly not, at least for examples as simple as this.
However, in terms of uniform approximations, there are advantages:
\vspace*{-0.2em}
\begin{enumerate}
 \item 
 The process by which composite solutions are obtained with the approach
 in \S~\ref{sec:standard-matching} is not entirely simple --- particularly
 at higher orders. On the other hand, the approach in
 \S~\ref{SEIM::SBVP:sss:EqMa} yields uniform approximations directly.
 But this ignores the fact that solving the equations level matching
 equations is, generally, harder than solving the ones from standard
 matching.
 \item \vspace*{-0.2em} 
 The uniform approximations produced by the equation level matching
 approach tend to be ``better'', in the sense that they are not only more
 accurate, but remain valid for a larger range in the small parameter.
 This is discussed in \S~\ref{sec:comp-betw-meth} for the example here,
 and later in \S~\ref{sec:example-from-acoust} for the acoustics example.
\end{enumerate}
Of course, in the context of an example as simple as this it is hard to
make meaningful comparisons. For this example writing the exact
solution 
handily beats both techniques. The meaningful differences arise
for wave problems, when the 
target is not a specific solution, but a simplified model equation for
the physics.
%
\subsection{Comparison between methods}\label{sec:comp-betw-meth}
Since an analytical solution to \eqref{eq:simple-bvp} exists, it is
useful to compare the approximations and their errors. First:
the exact solution to \eqref{eq:simple-bvp} is \vspace*{-0.2em}
\begin{align} \label{eq:simple-bvp-solution}
 u = (e^{\lambda_1\,x}-e^{\lambda_2\,(x-1)+\lambda_1})/(1-e^{\lambda_1 - \lambda_2})\/,
\end{align}
where $\lambda_j\/$ solves $\epsilon\,\lambda^2-\lambda+(1+\epsilon)=0\/$.
Note that $\lambda_2 = 1/\epsilon - \lambda_1\/$, where
$\lambda_1 = 1 + 2\,\epsilon + 4\,\epsilon^2\/ + \mathcal{O}(\epsilon^3)\/$
for $\epsilon \ll 1\/$.
Second, the standard matched asymptotic composite solution is, to leading
order \vspace*{-0.2em}
\begin{align} \label{eq:composite-expansions-solution}
  u = e^x - e^{(x-1)/\epsilon} + \mathcal{O}(\epsilon).
\end{align}
Finally, the leading order for the equation level matching
gives \vspace*{-0.1em}
\begin{align} \label{eq:ELM-composite-expansion}
 u_0 = (e^{\mu_1\,x} - e^{\mu_2\,(x-1)+\mu_1})/(1 - e^{\mu_1-\mu_2})\/,
\end{align}
where $\mu_j\/$ solves $\epsilon\,\mu^2-\mu+1=0\/$. It is easy to show
that the characteristic values $\mu\/$ and $\lambda\/$ are related by
$\lambda_1 = \mu_1 + \nu\/$ and $\lambda_2 = \mu_2 -\nu\/$, where
$\nu = \epsilon + 2 \epsilon^2 + \dots$. Hence \emph{the characteristic
values for ${\mathcal L}\/$} (the $\mu_j\/$) \emph{approximate those of
the full problem up to errors that are small}
--- this is why $u_0\/$ can provide a uniform approximation.%

\begin{figure}[htb!]
\vspace*{-0.3em}\hspace*{0em}\hfill
\includegraphics[width=0.43\textwidth]{%
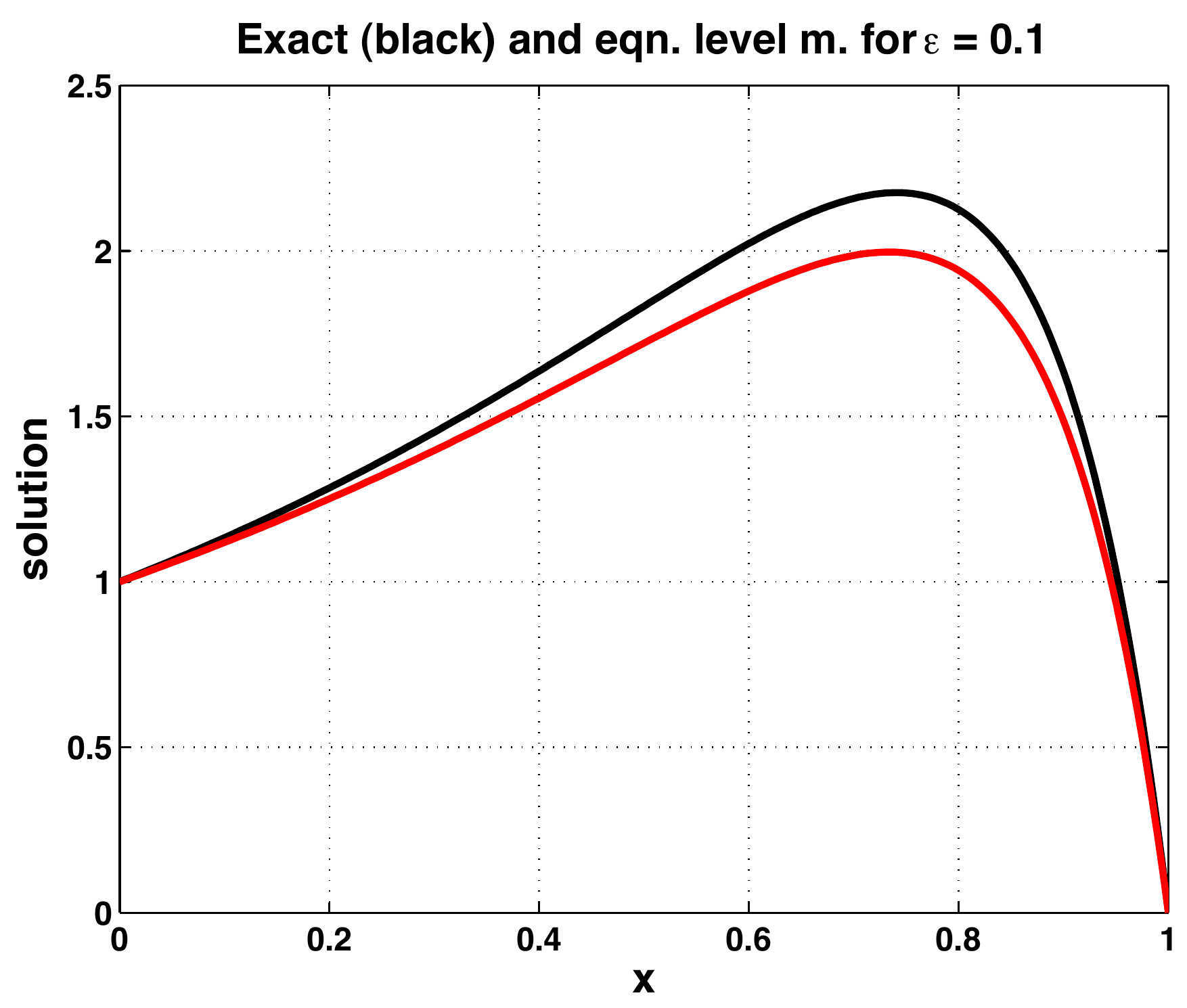} \hspace*{0.5em}
\includegraphics[width=0.43\textwidth]{%
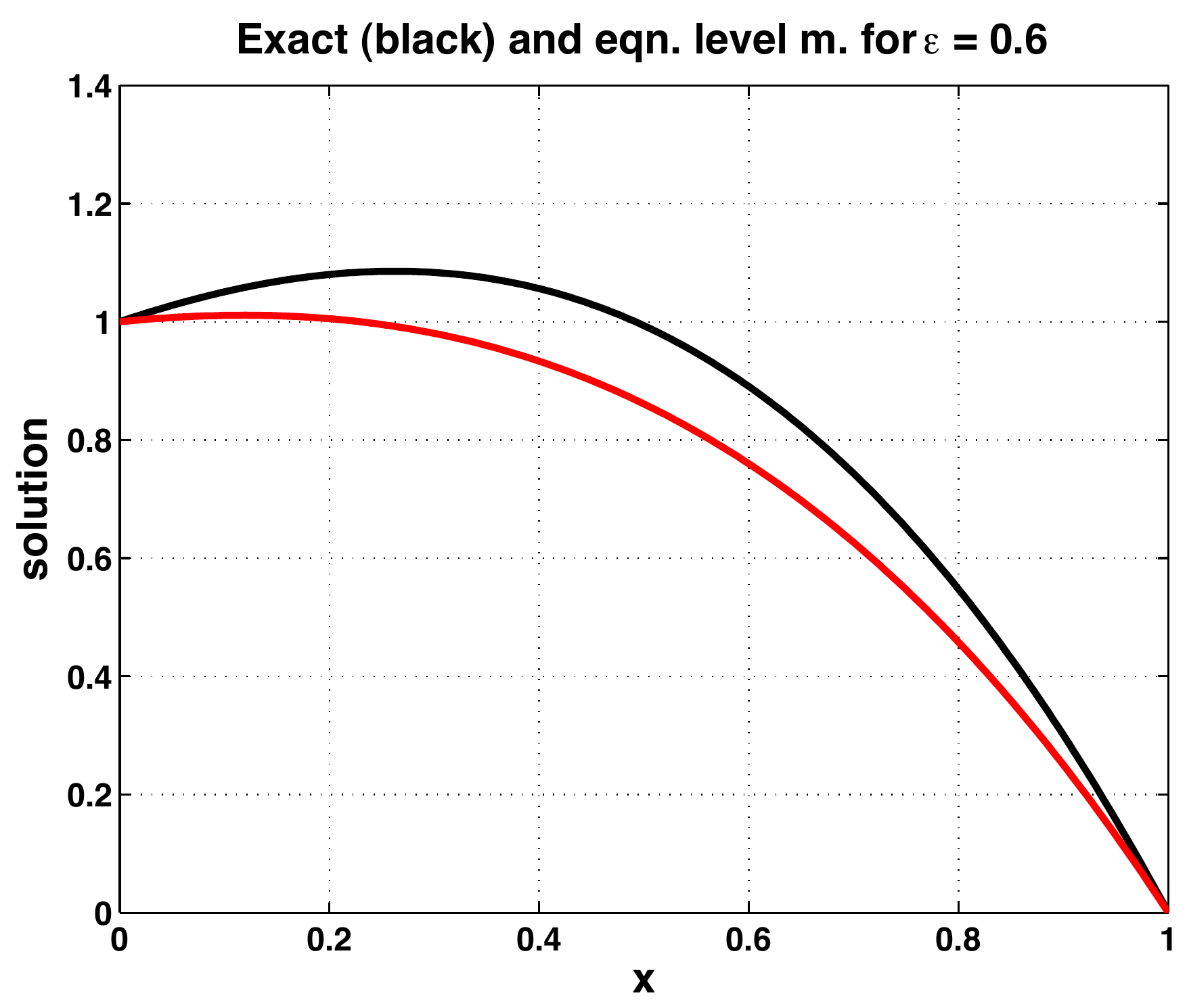}\hfill\hspace*{0em}
\vspace*{-0.7em}

\hfill\parbox[b]{1\textwidth}{\caption[Comparisons of the exact
 versus the equation level matching approximation.]{\small\sf
 BVP in \eqref{eq:simple-bvp}. Comparison: exact solution (black)
 versus equation level matching first term (red). Left: $\epsilon=0.1\/$.
 Right: $\epsilon=0.6\/$, showing that, even for ``large'' values of
 $\epsilon\/$, equation level matching provides a qualitative correct
 (even if not very accurate) approximation.
}\label{SEIM:fig:01}}\hfill\vspace*{-0.4em}
\end{figure}

\begin{figure}[htb!]
\vspace*{-0.3em}\hspace*{0em}\hfill
\includegraphics[width=0.43\textwidth]{%
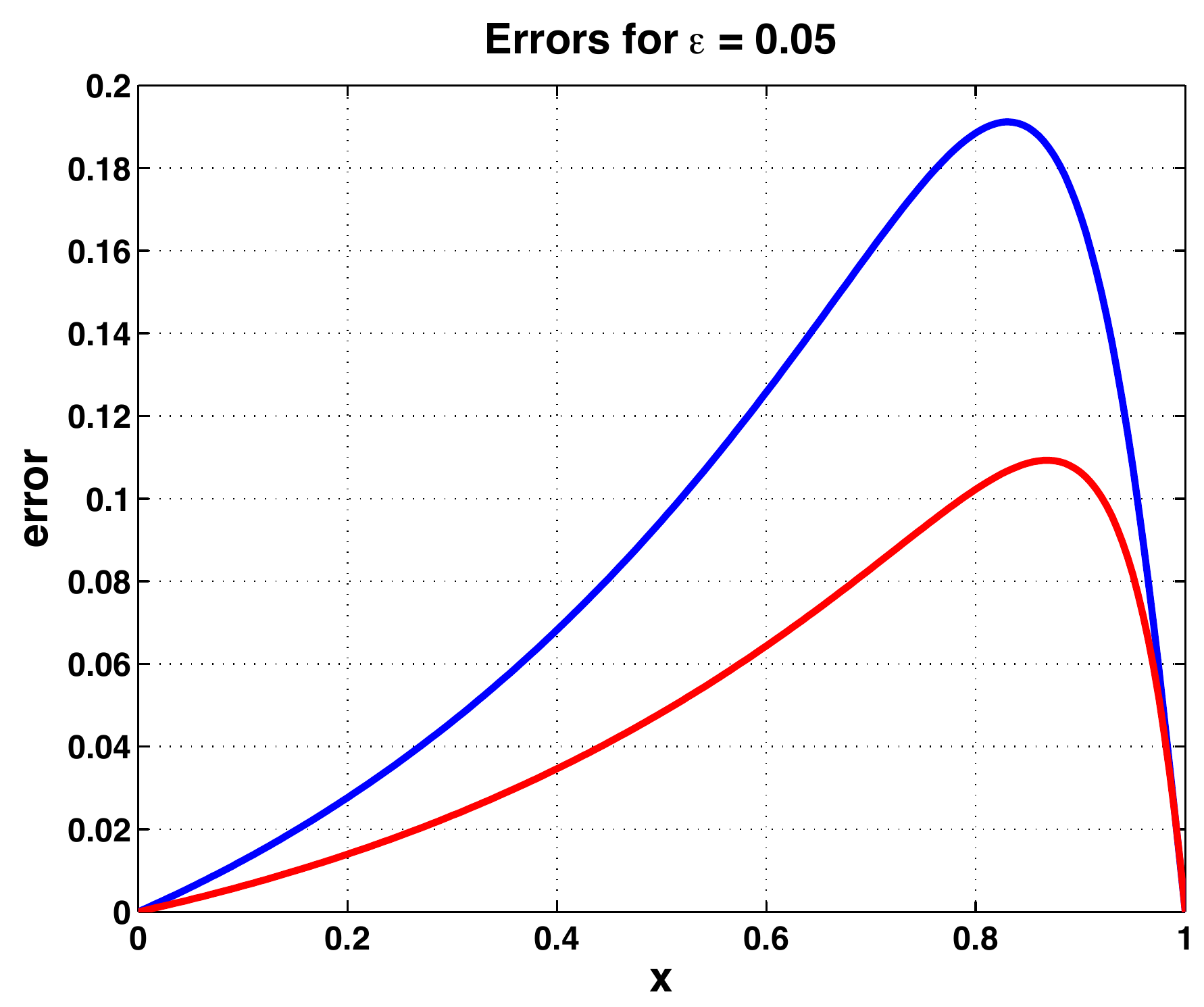} \hspace*{0.5em}
\includegraphics[width=0.43\textwidth]{%
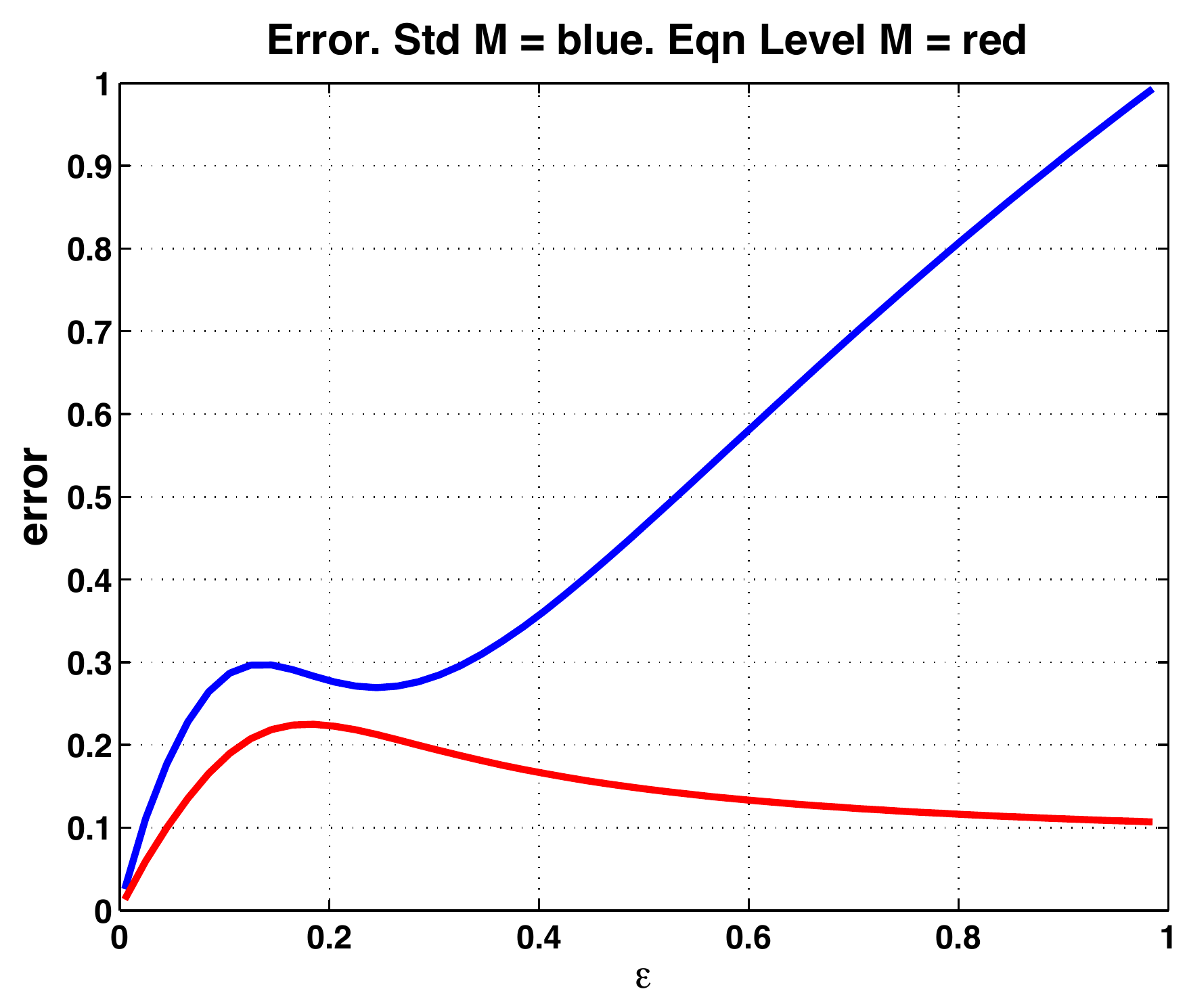}\hfill\hspace*{0em}
\vspace*{-0.7em}

\hfill\parbox[b]{1\textwidth}{\caption[Error plots for matching:
 standard versus equation level]{\small\sf BVP in \eqref{eq:simple-bvp}.
 Errors for the approximate solutions. Blue: leading order
 standard matching composite solution. Red: equation level matching first
 term. Left: error, as a function of $x\/$, for $\epsilon=0.05\/$. Right:
 $L_\infty\/$ norm of the error, as a function of $\epsilon\/$. The
 equation level matching error remains fairly small for all values of
 $\epsilon\/$ --- see text for more details.
}\label{SEIM:fig:02}}\hfill\vspace*{-0.4em}
\end{figure}

Note that \vspace{-0.2em}
\begin{enumerate}
 \item 
 (\ref{eq:ELM-composite-expansion}) is substantially more complicated than
 \eqref{eq:composite-expansions-solution}, but not alarmingly so. The
 extra complication is compensated by increased accuracy, and (more
 important) qualitative validity even for not small $\epsilon$-values.
 See figures~\ref{SEIM:fig:01}--\ref{SEIM:fig:02}.
 \item \vspace*{-0.2em} 
 The higher order terms grow rapidly in complication. However, if the
 objective is to capture the essential behavior at leading order (often
 the case in wave problems), higher order terms are not important.
\end{enumerate}
%
%
A comparison between the exact solution (\ref{eq:simple-bvp-solution}),
and \eqref{eq:ELM-composite-expansion} is show in
figure~\ref{SEIM:fig:01} for two different values of $\epsilon\/$. An
important feature is that, besides being a uniformly valid approximation
for $\epsilon\/$ small, \emph{it behaves qualitatively correct even for
large values of $\epsilon\/$.}
%
%
Validity for a wide $\epsilon$-range is an attractive property for complex
physical situations, where the aim is to obtain as simple a model as
possible, for the purpose of understanding the observed behaviors ---
often the case in wave research.

%
%
Figure~\ref{SEIM:fig:02} shows that equation level matching leads to
errors which are both smaller, and remain small for large $\epsilon\/$.
Note: that the error remains small even for huge $\epsilon$-values (it
keeps decreasing beyond $\epsilon=1\/$, and asymptotes
$\cot(1) - \arccos(\sin(1)) \approx 0.071\/$); this is a consequence
of the simplicity of the example, and should not be taken seriously
%

\section{Example from waves theory: weakly nonlinear
  detonations}\label{sec:SWND}
In this example we use the method to study weakly nonlinear waves in a
reactive gas. The goal is to rationally treat dissipation as a singular
perturbation problem, and derive the appropriate model incorporating
dissipative effects. For reactive gas dynamics background information see,
e.g., \cite{fickett2011detonation,williams1985combustion,Lee-2008}. The
main technical issue to address is: in the limit of high activation energy
and weak heat release, kinetic theory predicts exponentially small
diffusion coefficients.\,\footnote{\sf \,For inert gases it is
   possible to consider length scales where dissipative effects can
   be incorporated via a weakly nonlinear wave expansion --- e.g.,
   derivation of Burgers' equation. But chemical reactions introduce a
   spacial scale, the reaction length, which is much larger than the
   scale where dissipation plays a role.}
Hence in weakly nonlinear wave perturbation theory these effects are left
out~\cite{rosales1989diffraction,clavin2002dynamics,faria2015theory}.
However, there is some current interest on the influence of transport
effects (viscosity, thermal conductivity, and species diffusion) on the
stability of detonations~\cite{barker2015viscous,romick2015verified,%
romick2012effect}. Thus an investigation of these effects within a
weakly nonlinear model seems appropriate.
%
\subsection{The mathematical model}
Here we will consider an abstract version of the 1-D reacting equations
for compressible gas dynamics. Specifically, in adimensional
variables: \vspace*{-0.2em}
\begin{eqnarray}
 \vec{u}_t + (\vec{F}(\vec{u}))_x & = & \epsilon^2\,\vec{W} \left(
    \tfrac{1}{\epsilon}\,\vec{u}\/,\,\lambda\right) + \epsilon\,\delta
    \,\mathcal{D}(\partial_x)\,\vec{u}\/, \label{SWND:eqn:01}\\
 \lambda_t + \phi\left(\vec{u}\right)\,\lambda_x & = & w \left(
    \tfrac{1}{\epsilon}\,\vec{u}\/,\,\lambda\right)\/,\label{SWND:eqn:02}
\end{eqnarray}
where $\vec{u}\/$ is the vector of perturbations (from a constant rest
state) to the ``fluid'' conserved densities, $\vec{F}\/$ is the vector of
fluxes, $0\leq\lambda\leq 1\/$ is a reaction progress variable,
$\phi\/$ is the ``particle speed'', $0 < \epsilon \ll 1\/$ is the inverse
of the ``activation energy'', and $\delta>0\/$ is a parameter
characterizing the transport effects --- assumed exponentially small in
$\epsilon\/$. $\vec{W}\/$ is the vector of sources caused by the
``reaction'' (the prefactor $\epsilon^2\/$ indicates ``weak heat
release''), while $w \geq 0\/$ is the reaction rate. Finally,
$\mathcal{D}(\partial_x)=D\,\partial_x^2\/$ ($D\/$ a constant
square matrix) is a diffusion operator: all the solutions to the
linearized ($\vec{u}\/$ infinitesimal), non-reacting
(no $\lambda\/$), problem decay in time.

We assume that $\vec{F}\/$ is smooth, and that the square matrix
$\nabla (\vec{F})\/$ has distinct real eigenvalues (\emph{the system is
strictly hyperbolic}). Expanding for $\vec{u}\/$ small
\vspace*{-0.2em}
\begin{equation} \label{SWND:eqn:03}
 \vec{F} = \vec{F}_0 + A_1\,\vec{u} + \vec{A}_2 (\vec{u}\/,\,\vec{u})
       + \vec{A}_3 (\vec{u}\/,\,\vec{u}\/,\,\vec{u}) + \dots\/,
       \quad A_1 = \nabla (\vec{F})(\vec{0})\/,
         \vspace*{-0.3em}
\end{equation}
where $\vec{A}_2$ is a symmetric vector valued bilinear function, etc.

The motivation for the assumed scalings is: we seek for a situation where
the waves are weakly nonlinear. Because of the reaction terms, such waves
are possible only for a suitably small heat release. Further, for the weak
waves to couple with the reaction, sensitive dependence of the
reaction terms on the fluid variables is needed, as provided by the
high activation energy assumption.
%
\subsection{Weakly nonlinear detonation waves expansion}
\label{SWND:sub:WNDW}
Let $c\/$ be an eigenvalue of $A_1\/$ such that
$c \neq \phi_0 = \phi(\vec{0})\/$. Let $\vec{l}\/$ and $\vec{r}\/$ be
left and right eigenvectors of $A_1\/$ corresponding to $c\/$, normalized
by $\vec{l} \cdot \vec{r} = 1\/$. Then we propose a weakly nonlinear
traveling wave expansion:
\begin{equation} \label{SWND:eqn:04}
 \vec{u} \sim \epsilon\,\vec{u}_1(\chi\/,\,\tau)
            + \epsilon^2\,\vec{u}_2(\chi\/,\,\tau) + \dots
 \quad \mbox{and} \quad \lambda \sim \lambda_0(\chi\/,\,\tau) +
            \epsilon\,\lambda_1(\chi\/,\,\tau) + \dots\/,
\end{equation}
where $\chi = x-c\,t\/$ and $\tau = \epsilon\,t\/$. Assume that the wave
moves into the rest state ($\vec{u}\/$ vanishes for $\chi \to \infty$).
Substituting into (\ref{SWND:eqn:01}), and collecting equal powers
of $\epsilon$,\vspace*{-0.2em}
\begin{eqnarray}
 {\mathcal O}(\epsilon^1) \;\; (A_1-c)\,(\vec{u}_1)_\chi &=&
 0\/.  \label{SWND:eqn:05} \\
 {\mathcal O}(\epsilon^2) \;\; (A_1-c)\,(\vec{u}_2)_\chi &=&
 - (\vec{u}_1)_\tau - (\vec{A}_2(\vec{u}_1\/,\,\vec{u}_1))_\chi
 + \vec{W}(\vec{u}_1\/,\,\lambda_0)\/.\;\;\; \label{SWND:eqn:06}\\
 \dots &=& \dots \nonumber\\
 {\mathcal O}(\epsilon^n) \;\; (A_1-c)\,(\vec{u}_n)_\chi &=&
 - (\vec{u}_{n-1})_\tau - (\vec{N}_n)_\chi + \vec{W}_{n-2}\/.
 \label{SWND:eqn:07}
\end{eqnarray} 
In (\ref{SWND:eqn:07}):\hspace*{0.6em}
(i)   $n \geq 2\/$;\hspace*{0.6em}
(ii)  $\vec{N}_n = \sum_{j \geq 2} \;\; \sum_{n_1+\dots+n_j=n}
         \vec{A}_j\left(\vec{u}_{n_1}\/,\, \dots \vec{u}_{n_j}\right)\/$,
      \;with $n_q \geq 1\/$;\hspace*{0.6em}
(iii) $\vec{W}_q\/$ is the $q$-th term in the expansion of $\vec{W}\/$
      --- note that $\vec{W}_q$ depends on $\vec{u}_1\/,\, \dots
         \vec{u}_{q+1}\/$ and $\lambda_0\/,\,\dots \lambda_q\/$ only.
Because of the assumption on the size of $\delta\/$, no dissipative
terms appears at any order.

The $\mathcal{O}(\epsilon)\/$ equations yield \vspace*{-0.4em}
\begin{align}
  \vec{u}_1 = \sigma_1\,\vec{r}\/,
\end{align}
where $\sigma_1 = \sigma_1(\chi\/,\,\tau)\/$ is scalar valued. For the
$\mathcal O(\epsilon^2)$ equations we write
$\vec{u}_2 = \sigma_2\,\vec{r} + \vec{v}_2$, where $\vec{v}_2$ is a linear
combination of the right eigenvectors of $\vec{A}_1$ corresponding to
eigenvalues different from $c\/$ (thus $\vec{l} \cdot \vec{v}_2 = 0\/$).
Then \eqref{SWND:eqn:06} has a solution if and only if the right hand
side is orthogonal to $\vec{l}$. Thus \vspace*{-0.2em}
\begin{equation} \label{SWND:eqn:10}
 (\sigma_1)_\tau + (a\,\sigma_1^2)_\chi =
 \vec{l}\cdot\vec{W}\left(\sigma_1\,\vec{r}\/,\,\lambda_0\right)\/,
\end{equation}
where $a = \vec{l} \cdot \vec{A}_2(\vec{r}\/,\,\vec{r})\/$. In general,
at any order $\vec{u}_n = \sigma_n\,\vec{r} + \vec{v}_n\/$, with
$\vec{l} \cdot \vec{v}_n  = 0\/$. Then $\vec{v}_n\/$ follows from
\eqref{SWND:eqn:07}, and $\sigma_n\/$ from the solvability condition at
the next order --- the right had side of \eqref{SWND:eqn:07} must be
orthogonal to $\vec{l}\/$ at all orders
(see \S~\ref{SWND:WNDW:sss:THOE}).

Finally the expansion for $\lambda\/$ is governed by \vspace*{-0.2em}
\begin{eqnarray}
 {\mathcal O}(\epsilon^0) \;\; (\phi_0-c)\,(\lambda_0)_\chi &=&
  w\,\left(\vec{u}_1\/,\,\lambda_0\right)\/, \label{SWND:eqn:12}\\
 \dots &=& \dots \nonumber \\
 {\mathcal O}(\epsilon^n) \;\; (\phi_0-c)\,(\lambda_n)_\chi &=&
 - (\lambda_{n-1})_\tau -\mbox{$\sum_{j=1}^n$}\,\phi_j
   \,(\lambda_{n-j})_\chi + w_n\/,\;\; \label{SWND:eqn:13}
\end{eqnarray}
where the BC $\lambda = 0\/$ is imposed for $\chi \to \infty\/$ (ahead
of the wave). In (\ref{SWND:eqn:13}) $\phi_q\/$ and $w_q\/$ are the
$q$-th terms in the expansions for $\phi\/$ and $w\/$. Note that $w_n\/$
involves dependence on $\vec{u}_{n+1}\/$ and $\lambda_n\/$, via the terms
$(\vec{u}_{n+1} \cdot \nabla_u)\,w(\vec{u}_1\/,\,\lambda_0)$ and
$\lambda_n\,w_\lambda(\vec{u}_1\/,\,\lambda_0)\/$.
%
\subsubsection{The higher orders}\label{SWND:WNDW:sss:THOE}
The leading order equations controlling the expansion are
(\ref{SWND:eqn:10}) and (\ref{SWND:eqn:12}).  Next we describe the higher
order equations. Substitute $\sigma_1 \to \sigma_1 + \Delta\sigma\/$ and
$\lambda_0 \to \lambda_0 + \Delta\lambda\/$ into (\ref{SWND:eqn:10})
and (\ref{SWND:eqn:12}), where $\Delta\sigma\/$ and
$\Delta\lambda\/$ are infinitesimals. Then $\Delta\sigma\/$ and
$\Delta\lambda\/$ satisfy linear homogeneous equations (with coefficients
that depend on $\sigma_1\/$ and $\lambda_0\/$)  \vspace*{-0.2em}
\begin{align*}
 {\mathcal L}\,\vec{\Delta}=\vec{0}\/,
\end{align*}
where $\vec{\Delta}\/$ is the vector with components $\Delta\sigma\/$ and
$\Delta\lambda\/$. The higher order equations are forced versions of this
equation: ${\mathcal L}\,\vec{\Delta}_n=\vec{f}_n\/$, $n \geq 2\/$, where
$\vec{f}_n\/$ depends on the lower order terms only, and $\vec{\Delta}_n$
is the vector with components $\sigma_n\/$ and $\lambda_{n-1}\/$.
In specific examples one can see that these equations can develop
secularities for $\chi\/$ or $\tau\/$ large. \emph{The expansion in this
section requires $\chi = o(1/\epsilon)\/$ and $\tau = o(1/\epsilon)\/$.}
%
\subsection{Burgers' shock waves expansion}\label{SWND:sub:BSWE}
Change variables in (\ref{SWND:eqn:01}--\ref{SWND:eqn:02}), to
$T = t/\delta\/$ and $X = x/\delta\/$. Then\vspace*{-0.2em}
\begin{eqnarray}
 \vec{u}_T + (\vec{F} (\vec{u}))_X & = & \epsilon^2\,\delta\,\vec{W}
    (\tfrac{1}{\epsilon}\,\vec{u}\/,\,\lambda) +
    \epsilon\,{\mathcal D}(\partial_X)\,\vec{u}\/, \label{SWND:eqn:14}\\
 \lambda_T + \phi (\vec{u})\,\lambda_X & = & \delta\,w
    (\tfrac{1}{\epsilon}\,\vec{u}\/,\,\lambda)\/. \label{SWND:eqn:15}
\end{eqnarray}
Expand\vspace*{-0.2em}  
\begin{eqnarray}
 \vec u &\sim& \epsilon\,\vec u_1(\xi\/,\,\mu)
               + \epsilon^2\,\vec u_2(\xi\/,\,\mu) + \dots \nonumber\\
 \mbox{and}\;\;
 \lambda &\sim& \lambda_0(\xi\/,\,\mu) +
            \epsilon\,\lambda_1(\xi\/,\,\mu) + \dots\/,
 \label{SWND:eqn:16}
\end{eqnarray}
where $\xi = X-c\,T\/$ and $\mu = \epsilon\,T\/$. Then\vspace*{-0.2em}
\begin{eqnarray}
 {\mathcal O}(\epsilon^1) \;\; (A_1-c)\,(\vec u_1)_\xi &=&
                                    0\/.  \label{SWND:eqn:17}\\
 {\mathcal O}(\epsilon^2) \;\; (A_1-c)\,(\vec u_2)_\xi &=&
 - (\vec u_1)_\mu - (\vec{A}_2(\vec u_1\/,\,\vec u_1))_\xi +
 \mathcal D(\partial_\xi)\,\vec u_1\/.\;\;  \label{SWND:eqn:18}\\
 \dots &=& \dots \nonumber\\
 {\mathcal O}(\epsilon^n) \;\; (A_1-c)\,(\vec u_n)_\xi &=&
 - (\vec u_{n-1})_\mu - (\vec{N}_n)_\xi +
 \mathcal D(\partial_\xi)\,\vec u_{n-1}\/.\label{SWND:eqn:19}
\end{eqnarray}
The solution to these equations follows the same pattern as for
(\ref{SWND:eqn:05}--\ref{SWND:eqn:07}). Write\vspace*{-0.2em}
\begin{equation} \label{SWND:eqn:20}
 \vec u_1 = \sigma_1\,\vec r \qquad\mbox{and}\qquad
 \vec u_n = \sigma_n\,\vec r + \vec v_n \;\;\mbox{for}\;\; n \geq 2\/,
\end{equation}
with $\vec l \cdot \vec v_n=0\/$. Then the $\sigma_n\/$ are determined
by the \emph{solvability conditions: the right hand sides in the equations
above must be orthogonal to} $\vec l\/$. In particular:\vspace*{-0.0em}
\begin{equation} \label{SWND:eqn:21}
 (\sigma_1)_\mu + (a\,\sigma_1^2)_\xi
   = \vec l \cdot \mathcal D(\partial_\xi)\,\vec r\,\sigma_1
   = \nu\,(\sigma_1)_{\xi\/\xi}\/,
\end{equation}
where $a\/$ is as in (\ref{SWND:eqn:10}) and
$\nu = \vec l\cdot D\,\vec r > 0\/$ is a constant --- $\nu > 0\/$
follows because $D\/$ is a diffusion matrix.
Note that (\ref{SWND:eqn:21}) \emph{is Burgers' equation}, the canonical
equation describing viscous weak shocks~\cite{burgers1948mathematical,%
whitham1974linear,choquet1968ondes,hunter1995asymptotic,%
rosales1991introduction}.

Similarly\vspace*{-0.2em}
\begin{eqnarray}
 \mathcal O(\epsilon^0) \qquad (\phi_0-c)\,(\lambda_0)_\xi &=&
   0\/. \label{SWND:eqn:22}\\
 \dots &=& \dots \nonumber \\
 \mathcal O(\epsilon^n) \qquad (\phi_0-c)\,(\lambda_n)_\xi &=&
   - (\lambda_{n-1})_\mu -
   \mbox{$\sum_{j=1}^n$}\,\phi_j\,(\lambda_{n-j})_\xi\/. \label{SWND:eqn:23}
\end{eqnarray}
These equations mean that no reaction occurs in this limit.
%
\subsection{Equation level matching} \label{SWND:sub:ELMA}
Next we extend the process introduced in
\S~\ref{sec:asympt-match-equat} to this section's
problem. Consider:\vspace*{-0.4em}
\begin{itemize}
 \item[(a)]
 The expansion in \S~\ref{SWND:sub:WNDW}, given by equations
    (\ref{SWND:eqn:05}--\ref{SWND:eqn:07}) and
    (\ref{SWND:eqn:12}--\ref{SWND:eqn:13}).
 \item[(b)] \vspace*{0.3em}
 The expansion in \S~\ref{SWND:sub:BSWE}, given by equations
    (\ref{SWND:eqn:17}--\ref{SWND:eqn:19}) and
    (\ref{SWND:eqn:22}--\ref{SWND:eqn:23}).\vspace*{-0.3em}
\end{itemize}
Then take the ``union'' of these expansions to produce a unified expansion
reducing to those in \S~\ref{SWND:sub:WNDW} and \S~\ref{SWND:sub:BSWE} in
the appropriate set of variables. Thus matching occurs at the equation
level. Hence, because the equations match, we know that the solutions also
match (in the sense of standard matched asymptotic expansions), without
the need to actually compute (or even know) the solutions.

We propose the unified expansion \vspace*{-0.2em}
\begin{eqnarray}
 \vec u & \sim & \epsilon\,\underbrace{\left(
       \sigma_1(\chi\/,\,\tau\/;\,\delta)\,\vec r\right)}_{\vec u_1} +
       \epsilon^2\,\underbrace{\left(
       \sigma_2(\chi\/,\,\tau\/;\,\delta)\,\vec r +
       \vec v_2(\chi\/,\,\tau\/;\,\delta)\right)}_{\vec u_2} + \dots
       \label{SWND:eqn:24}\\
 \lambda & \sim & \lambda_0(\chi\/,\,\tau\/;\,\delta) +
            \epsilon\,\lambda_1(\chi\/,\,\tau\/;\,\delta) + \dots\/,
       \label{SWND:eqn:25}
\end{eqnarray}
where $\vec l \cdot \vec v_n = 0\/$. \emph{Here, unlike
(\ref{SWND:eqn:04}) or (\ref{SWND:eqn:16}), the terms include an explicit
dependence on the small parameter $\delta\/$. A term by term formal
expansion is not possible. The equations must be obtained via matching,
as follows:} \vspace*{-0.2em}
\begin{eqnarray}
 (A_1-c)\,(\vec u_1)_\chi &=& 0\/, \label{SWND:eqn:26}\\
 (A_1-c)\,(\vec u_2)_\chi &=& - (\vec u_1)_\tau -
    (\vec{A}_2(\vec u_1\/,\,\vec u_1))_\chi \nonumber\\ &&
    +\vec W(\vec u_1\/,\,\lambda_0)  +
    \delta\,\mathcal D(\partial_\chi)\,\vec u_1\/, \label{SWND:eqn:27}\\
 \dots &=& \dots \nonumber\\
 (A_1-c)\,(\vec u_n)_\chi &=& - (\vec u_{n-1})_\tau - (\vec{N}_n)_\chi +
    \vec W_{n-2} + \delta\,\mathcal D(\partial_\chi)\,\vec u_{n-1}\/,
    \;\;\label{SWND:eqn:28}
\end{eqnarray}
where we use the same notation as in \S~\ref{SWND:sub:WNDW} and
\S~\ref{SWND:sub:BSWE}. Similarly  \vspace*{-0.2em}
\begin{eqnarray}
 (\phi_0-c)\,(\lambda_0)_\chi &=& w\left(\vec u_1\/,\,\lambda_0\right)\/,
    \label{SWND:eqn:29}\\
 \dots &=& \dots \nonumber \\
 (\phi_0-c)\,(\lambda_n)_\chi &=& - (\lambda_{n-1})_\tau
    -\mbox{$\sum_{j=1}^n$}\,\phi_j\,(\lambda_{n-j})_\chi
    + w_n\/. \label{SWND:eqn:30}
\end{eqnarray}
This expansion matches the one in \S~\ref{SWND:sub:WNDW}
because\,\footnote{\sf \,Recall that $\delta$ is smaller than any
   power of $\epsilon\/$.}
the effect of the terms multiplied by $\delta\/$ in the equations occurs
beyond all the $\vec u_n\/$ and $\lambda_n\/$, so that the equations
reduce to those in (\ref{SWND:eqn:05}--\ref{SWND:eqn:07}) and
(\ref{SWND:eqn:12}--\ref{SWND:eqn:13}).
Rewriting the equations in terms of the variables
$\xi = \chi/\delta\/$ and $\mu = \tau/\delta\/$, it is easy to
see (same argument) that the expansion here matches the one in
\S~\ref{SWND:sub:BSWE} --- the equations reduce to those in
(\ref{SWND:eqn:17}--\ref{SWND:eqn:19}) and
(\ref{SWND:eqn:22}--\ref{SWND:eqn:23}).

Finally, note that \emph{the leading order uniform expansion is given by}
\begin{eqnarray}
 (\sigma_1)_\tau + (a\,\sigma_1^2)_\chi &=&
 \vec l\cdot\vec W\left(\sigma_1\,\vec r\/,\,\lambda_0\right) +
 \nu\,\delta\,(\sigma_1)_{\chi\/\chi}\/, \label{SWND:eqn:31}\\
 (\phi_0-c)\,(\lambda_0)_\chi &=&
 w\left(\vec u_1\/,\,\lambda_0\right)\/.
 \label{SWND:eqn:32}
\end{eqnarray}
The equations in~\cite{RosalesMajda1983wndw} reduce to the ones above
in the plane-wave case. However, in~\cite{RosalesMajda1983wndw} the
non-physical assumption $\delta \sim \epsilon$ was introduced. A similar
approach to incorporate transport effects in a more elaborate setting
(including 2-D effects, species diffusion and decoupling of the thermal
effects) can in be found in~\cite{faria2014thesis}.

The dissipative term in (\ref{SWND:eqn:31}--\ref{SWND:eqn:32}) adds
structure to the shocks, which in the inviscid case are just point
discontinuities moving in space-time. Although this is only of secondary
importance when studying the steady traveling wave solutions to
(\ref{SWND:eqn:31}--\ref{SWND:eqn:32}), the finite shock width appears to
play a more important role in the highly unstable regime where detonations
are typically found \cite{barker2015viscous,romick2015verified}. More
importantly, including dissipation in a modified version of the weakly
nonlinear theory presented in this paper can trigger new types of
subsonic traveling wave solution not present in the inviscid theory
\cite{fariaICDERS2015dissipative}. Further physically
relevant questions occur in relationship with the role of dissipation for
multi-dimensional wave interactions. This is the subject of current study
by the authors, and will be reported elsewhere.
\section{Example: dissipation from the acoustic boundary layer}
\label{sec:example-from-acoust}
One topic we hope this method is useful for is: incorporate into
wave models dissipation and/or dispersion, when these effects occur at
higher order in standard asymptotic wave theory.
In these situations conventional matching can fail as well because, for
example: (i) The behavior is too complex, and there are no solutions
available for the matching process. (ii) The time behavior for the
``inner'' and ``outer'' expansions differs. In this section we study
a simple example of the situation in (ii), and consider the effects of
boundary layer dissipation in acoustics. In \S~\ref{LAIC:sub:ELM1} we
include a brief description of the standard methods in this subject.

For simplicity, we work in 2-D, where the linearized
isentropic\,\footnote{\sf \,In particular this excludes the thermal
   boundary layer. This is meant as an illustrative example.}
Navier-Stokes equations for flow a channel, $0 < \tilde y < L\/$ and
$-\infty < \tilde x < \infty\/$ are \vspace*{-0.2em}
\begin{eqnarray}
 \tilde \rho_{\tilde t} + \rho_0\,\left(\tilde u_{\tilde x} +
 \tilde v_{\tilde y}\right) &=& 0\/,
    \label{LAIC:eqn:01}\\
 \tilde u_{\tilde t} + \tfrac{c_0^2}{\rho_0}\,\tilde \rho_{\tilde x} &=&
    \nu\,\tilde{\Delta}\,\tilde u\/,\label{LAIC:eqn:02}\\
 \tilde v_{\tilde t} + \tfrac{c_0^2}{\rho_0}\,\tilde \rho_{\tilde y} &=&
    \nu\,\tilde{\Delta}\,\tilde v\/.\label{LAIC:eqn:03}
\end{eqnarray}
Here $\rho_0 =$ ambient density, $c_0 =$ sound speed, and $\nu =$
kinematic viscosity. The tildes denote dimensional variables.
Nondimensionalize:
$\tilde \rho = \rho_0\,\rho\/$,
$\tilde u = c_0\,u\/$,
$\tilde v = \epsilon\,c_0\,v\/$,
$\tilde x = \lambda\,x\/$,
$\tilde y = L\,y\/$, and
$\tilde t = (\lambda/c_0)\,t\/$,
where $\lambda\/$ is a typical wavelength and $\epsilon = L/\lambda\/$.
Then \vspace*{-0.2em}
\begin{align} \label{eq:acoustic-nondim-1}
  \rho_t + u_x + v_y &= 0\/,\\ \label{eq:acoustic-nondim-2}
  u_t + \rho_x &= \tfrac{\delta}{\epsilon}\,\Delta\/u\/,\\
     \label{eq:acoustic-nondim-3}
  \epsilon^2\,v_t + \rho_y &= \delta\,\epsilon\,\Delta\/v\/,
\end{align}
where $\delta = \nu/(c_0\,L)\/$ and
$\Delta = \partial_y^2 + \epsilon^2\,\partial_x^2\/$. 
The equations apply for $-\infty<x<\infty\/$ and $0<y<1\/$,
with boundary conditions: $u = v = 0\/$ at both $y = 0\/,\,1\/$.

\subsection{Historical perspective}\label{LAIC:sub:ELM1}
Here we consider expansions for
(\ref{eq:acoustic-nondim-1}--\ref{eq:acoustic-nondim-3})
in the regime $0 < \delta \ll \epsilon \ll 1\/$, where
$\epsilon = L/\lambda\/$ is the long wave parameter. In fact, we will
assume that $\delta = O(\epsilon^m) = \mathcal C\,\epsilon^m\/$, where
$\mathcal C>0\/$ is a constant and $m = 2\,p+1 > 1\/$ is an odd
integer.\,\footnote{\sf \,This is to simplify the algebra. Note that
   for some wind instruments (e.g., flute) $p = 1\/$ is reasonable.}

It should be obvious that a ``regular'' (powers of $\epsilon\/$) or
``outer''
expansion for (\ref{eq:acoustic-nondim-1}--\ref{eq:acoustic-nondim-3})
is inviscid at leading order, with the effects of viscosity appearing only
as forcing terms at higher order --- in fact, causing secular
growth.\,\footnote{\sf \,An easy way to see why secularities arise, is
  to do a regular expansion for the equation
  $\eta_t+\eta_x=\epsilon\,\eta_{x\/x}\/$.}
One could, in principle, eliminate these secularities via a multiple
times expansion. But this would be pointless, as the total dissipation
is usually dominated by the boundary layer dissipation. On this last
point: because the regular expansion can satisfy the BC for $v\/$, but
not for $u\/$, a boundary layer expansion is needed to complete the
picture. Unfortunately, because the outer/regular expansion does not
incorporate any boundary layer effects, it has the ``wrong'' time
dependence ---
lacks the decay that dissipation in the boundary layer induces. Thus
the regular expansion cannot actually be made to match with the boundary
layer expansion in the traditional sense --- there are no intermediate
scaled variables in which ``inner'' and ``outer'' solutions can be
re-expanded, so that they match term by term.  The reason is that the
standard approach to matched asymptotics is biased towards steady state
solutions, and ``arbitrary'' time dependences do not fit within it too
well.
\emph{Various solutions to these issues have been developed}
(e.g., see~\cite{Blackstock2000}). A few examples are:\vspace*{-0.2em}
\begin{enumerate}
 \item
  Calculate the eigenmodes for the full 2-D or 3-D problem, including the
  boundary conditions at the walls. Then approximate the exact ``dispersion
  relation'' thus obtained using the small
  parameters~\cite{Kirchhoff1868,Weston1953tpsw}. This approach goes back
  to the beginnings of the subject (acoustics). It precedes the
  development of the mathematical theory of matched asymptotic
  expansions~\cite{Kaplun1954,Kaplun1957,KaplunsPapers1967,%
  KaplunLagerstrom1957}, or even the introduction of the concept of
  boundary layer by L. Prandtl. Two disadvantages are:
  (i)  Very labor intensive.
  (ii) \emph{No extension to nonlinear problems.}
 \item \vspace*{-0.2em}
  Separate time, and use matched asymptotic expansions to calculate the
  eigenfunctions and eigenvalues. A variation of the approach in item~1.
  It has the advantage of going directly for the desired simplified
  dispersion relation, avoiding some messy calculations. But it is
  still restricted to the linear problem.
 \item \vspace*{-0.2em}
  Calculate the boundary layer dissipation per unit area for a time
  harmonic field, space independent and with flow parallel to the
  boundary. Then use the result to correct the Helmholtz equation
  --- see art. 328
  in~\cite{LambHydroBook}. This is a clever shortcut for the
  process in item~1. But it is limited to single frequency waves, with
  no clear extension to nonlinear problems. It is also slightly
  inconsistent: because of the dissipation, the waves outside the layer
  are not exactly harmonic (unless an external forcing is applied).
  Thus one should actually compute the dissipation by the layer of a
  non-harmonic forcing.
 \item \vspace*{-0.2em}
  Chester~\cite{Chester1964roct} generalized the approach in item~3, and
  computed the dissipation produced by an arbitrary time dependent velocity
  field, using a Laplace Transform approach. This is then converted into a
  drag per unit length along the tube, and inputted into a derivation of
  the governing equations by using conservation principles
  --- assuming longitudinal dependence, only, in the solution. This
  approach is applicable to ``not too nonlinear'' situations --- since
  then the boundary layer dissipation should still be a linear process.
\end{enumerate}
The (slight generalization) of the method of matched asymptotic expansions
introduced here allows us to bypass work-arounds such as ones above. It
also produces a final result which applies for a larger set of
parameter regimes. 
In particular, we expect it to capture the \emph{transition from
thin-boundary layer to fully viscous, as the wave-length grows} (or the
tube diameter is reduced) --- this is work in progress. Of course, the
extended model includes (in the appropriate regime) the prior ones.
%
\subsubsection{Acoustic outer/regular expansion}\label{LAIC:ELM1:AORE}
Start with (\ref{eq:acoustic-nondim-1}--\ref{eq:acoustic-nondim-3}) and
substitute expansions of the form
$\rho = \sum_{n=0}^\infty \epsilon^n\,\tilde \rho^n$, etc. Then collect
equal powers of $\epsilon\/$ \vspace*{-0.2em}
\begin{eqnarray}
 \tilde \rho^n_t + \tilde u^n_x + \tilde v^n_y & = & 0\/,
    \label{LAIC:ELM1:eqn:07}\\
 \tilde u^n_t + \tilde \rho^n_x & = & \mathcal C\,\tilde u^{n+1-m}_{y\/y} +
                 \mathcal C\,\tilde u^{n-1-m}_{x\/x}\/,
    \label{LAIC:ELM1:eqn:08}\\
 \tilde \rho^n_y & = & \mathcal C\,\tilde v^{n-1-m}_{y\/y} +
                \mathcal C\,\tilde v^{n-3-m}_{x\/x} -
                \tilde v^{n-2}_t\/,\label{LAIC:ELM1:eqn:09}
\end{eqnarray}
where $\tilde \rho^n = \tilde u^n = \tilde v^n = 0\/$ if $n < 0\/$, and
$\tilde v_n = 0\/$ for $y=0\/$ or $y=1\/$ --- the BC for
$u\/$ cannot be satisfied. Note that: \vspace*{-0.2em}
\begin{enumerate}
 \item
 The leading order equations are
 $\tilde \rho^0_t+\tilde u^0_x+\tilde v^0_y =
        \tilde u^0_t+\tilde \rho^0_x=\tilde \rho^0_y = 0\/$.
 Average these equations over $y\/$, and use the BC to obtain the 1-D
 acoustic equations\;
 $\tilde \rho^0_t+\tilde u^M_x = \tilde u^M_t+\tilde \rho^0_x = 0\/$,\;
 where $\tilde u^M = \int_0^1 \tilde u^0 d\/y\/$.
 \item  \vspace*{-0.2em}
 The higher orders in this expansion develop \emph{secular
 behavior in time}, through resonances with the lower orders.
 These are related to \emph{bulk dissipation}, since this expansion ignores
 the boundary layers. 
 It may be possible to eliminate these secularities using
 multiple time scales. Here we will ignore them, as they occur on time
 scales much  longer than those associated with the boundary layer
 dissipation --- which we will incorporate, see \S~\ref{LAIC:ELM1:TUEX}.
\end{enumerate}
%
\subsubsection{The acoustic boundary layer expansion}
\label{LAIC:ELM1:ABLE}
The boundary layers at $y = 0\/,\,1\/$ can be treated in exactly the same
way, so we only show the calculations for $y = 0\/$. Start with
(\ref{eq:acoustic-nondim-1}--\ref{eq:acoustic-nondim-3}) and change
variables to $V = \epsilon\,v\/$ and $Y = (\sqrt{\epsilon/\delta})\,y\/$.
Then \vspace*{-0.2em}
\begin{equation}\label{LAIC:ELM1:eqn:10}
 \xi\,(\rho_t + u_x) + V_Y = 0\/, \quad
 u_t + \rho_x = \Delta_b\/u\/,\quad
 \xi\,V_t + \rho_Y = \xi\,\Delta\/V\/,
\end{equation}
where $\Delta_b = \partial_Y^2 + \xi^2\,\partial_x^2\/$ and
$\xi =\sqrt{\epsilon\,\delta} =\sqrt{\mathcal C}\,\epsilon^{p+1}\/$. Note
that when $p\/$ is not an integer, $\xi\/$ involves fractional powers.
This is not a problem, but it requires that extra terms be added to the
regular expansion, to make matching possible (a well known phenomenon in
matched asymptotic expansions). To gain simplicity, here we selected the
relationship between $\delta\/$ and $\epsilon\/$ to avoid the effect.

Now substitute into (\ref{LAIC:ELM1:eqn:10}) expansions of the form
$\rho = \sum_{n=0}^\infty \epsilon^n\,\bar \rho^{\,n}\/$, etc., which leads
to \vspace*{-0.2em}
\begin{eqnarray}
 \bar v^n_Y & = & -\sqrt{\mathcal C}\,\left(\bar \rho^{\;n-1-p}_t +
                 \bar u^{\;n-1-p}_x\right)\/,\label{LAIC:ELM1:eqn:11}\\
 \bar u^n_t + \bar \rho^n_x - \bar u^n_{Y\/Y} & = &
    \mathcal C\,\bar u^{\;n-2-2\,p}_{x\/x}\/, \label{LAIC:ELM1:eqn:12}\\
 \bar \rho^n_Y & = & \sqrt{\mathcal C}\,\bar v^{\;n-1-p}_{Y\/Y} +
         \mathcal C^{3/2}\,\bar v^{\;n-3-3\,p}_{x\/x} -
        \sqrt{\mathcal C}\,\bar v^{\;n-1-p}_t\/,\label{LAIC:ELM1:eqn:13}
\end{eqnarray}
where $\bar \rho^n = \bar u^n = \bar v^n = 0\/$ if $n < 0\/$, and
$\bar u^n=\bar v^n=0\/$ for $Y=0\/$.

These equations need BC for $Y\/$ large. In standard matching, these
would follow from matching with
(\ref{LAIC:ELM1:eqn:07}--\ref{LAIC:ELM1:eqn:09}). But this is impossible
because the expansions' time behaviors differ. Specifically:
(\ref{LAIC:ELM1:eqn:07}--\ref{LAIC:ELM1:eqn:09}) do not incorporate any
decay, while (\ref{LAIC:ELM1:eqn:11}--\ref{LAIC:ELM1:eqn:13}) do, via
$\bar u^n_{Y\/Y}\/$ in (\ref{LAIC:ELM1:eqn:12}). Though equation level
matching is possible, as shown next.
%
\subsubsection{The uniform expansion, expansion level matching}
\label{LAIC:ELM1:TUEX}
We use the process explained in \S~\ref{sec:asympt-match-equat} to
implement \eqref{pr:property-1}. Since the problem here is linear, modulo
technical details, the procedure is the same. Re-write
\eqref{eq:acoustic-nondim-1}--\eqref{eq:acoustic-nondim-3} in terms of
the linear operator $\vec{\mathcal L}\/$ which is the ``union'' of the
linear operators applied to the $n^{\rm th}\/$ order terms (on the left) in
(\ref{LAIC:ELM1:eqn:07}--\ref{LAIC:ELM1:eqn:09}),
(\ref{LAIC:ELM1:eqn:11}--\ref{LAIC:ELM1:eqn:13}), and the analog $y=1\/$.
That is: \vspace*{-0.2em}
\begin{eqnarray}
 \rho_t+ u_x + v_y \;=\; \mathcal L_1\,\vec u & = & 0\/,
    \label{LAIC:ELM1:eqn:14}\\
 u_t+ \rho_x - \tfrac{\delta}{\epsilon}\,u_{y\/y} \;=\;
    \mathcal L_2\,\vec u & = & \delta\,\epsilon\,u_{x\/x}\/,
    \label{LAIC:ELM1:eqn:15}\\
 \rho_y \;=\; \mathcal L_3\,\vec u & = & -\epsilon^2\,v_t+
    \delta\,\epsilon\,v_{y\/y} + \delta\,\epsilon^3\,v_{x\/x}\/,
    \label{LAIC:ELM1:eqn:16}
\end{eqnarray}
where $\vec u = (\rho\/,\,u\/,\,v)\/$ 
and the $\mathcal L_j\/$ are the components of $\vec{\mathcal L}\/$,
defined by the formulas. Then the \emph{equation level uniform expansion}
is $\vec u \sim \sum_{n=0}^\infty \,\epsilon^n\,\vec u^{\,n}\/$,
with \vspace*{-0.2em}
\begin{eqnarray}
 \mathcal L_1\,\vec u^{\,n} & = & 0\/, \label{LAIC:ELM1:eqn:17}\\
 \mathcal L_2\,\vec u^{\,n} & = & \mathcal C\,u^{n-1-m}_{x\/x}\/,
    \label{LAIC:ELM1:eqn:18}\\
 \mathcal L_3\,\vec u^{\,n} & = & \mathcal C\,v^{n-1-m}_{y\/y} +
    \mathcal C\,v^{n-3-m}_{x\/x} -v^{n-2}_t\/, \label{LAIC:ELM1:eqn:19}
\end{eqnarray}
where $\vec u^n = 0\/$ if $n < 0\/$, and $u^n = v^n = 0\/$ at both
$y=0\/,\,1\/$. Showing that
(\ref{LAIC:ELM1:eqn:07}--\ref{LAIC:ELM1:eqn:09}),
(\ref{LAIC:ELM1:eqn:11}--\ref{LAIC:ELM1:eqn:13}), and the analog for the
$y=1\/$ boundary, match [in the sense of \eqref{pr:property-1}]
via (\ref{LAIC:ELM1:eqn:17}--\ref{LAIC:ELM1:eqn:19}), can now be done
exactly as in \S~\ref{SEIM::SBVP:sss:EqMa}. The calculations are a bit
more cumbersome, but this is the only difference.

Let us now examine \emph{the leading order uniform equations}
\vspace*{-0.2em}
\begin{equation} \label{LAIC:ELM1:eqn:20}
 \rho_t + u_x + v_y = 0\/,\qquad
 u_t + \rho_x = \tfrac{\delta}{\epsilon}\,u_{y\/y}\/,
 \quad\mbox{and}\quad \rho_y = 0\/,
\end{equation}
where $u = v = 0\/$ at both $y=0\/,\,1\/$, and we have dropped the
superscript $0\/$ for notational simplicity. We can \emph{eliminate}
$v\/$, by averaging the first equation and using the BC \vspace*{-0.2em}
\begin{equation} \label{LAIC:ELM1:eqn:21}
 \rho_t + u^M_x = 0\/,\qquad
 u_t + \rho_x = \tfrac{\delta}{\epsilon}\,u_{y\/y}\/,
 \qquad u^M = \mbox{$\int$} u\,d\/y\/.
\end{equation}
Here both $\rho\/$ and $u^M\/$ are functions of $x\/$ and $t\/$ only.
We can go a bit further, by taking the average of the second equation as
well, \vspace*{-0.2em}
\begin{equation} \label{LAIC:ELM1:eqn:22}
 \rho_t + u^M_x = 0\/,\qquad
 u^M_t + \rho_x = \tfrac{\delta}{\epsilon}\,[u_{y}]\/,
\end{equation}
where $[u_{y}]\/$ denotes the jump in $u_y\/$ from $y=0\/$ to $y=1\/$. The
equations above are the \emph{equations for acoustic waves in a channel,
with boundary layer dissipation} incorporated. Just as with the example in
\S~\ref{sec:asympt-match-equat}, it \emph{appears as if the validity of
(\ref{LAIC:ELM1:eqn:20}) extends beyond its original intended range.} For
example, it encompasses compressible Poiseuille Flow, given by
\vspace*{-0.2em}
\begin{equation} \label{LAIC:ELM1:eqn:23}
 \rho = 2\,\tfrac{\delta}{\epsilon}\,x + \mbox{constant}\/, \quad
 u = y\,(1-y)\/, \quad\mbox{and}\quad v = 0\/,
\end{equation}
or its multiples. Hence (\ref{LAIC:ELM1:eqn:20}) may cover the whole
range, from nearly inviscid waves, to heavily dissipated waves, to the
transition where there are no more waves. We leave the investigation of
these question for another publication.

Of course for \eqref{LAIC:ELM1:eqn:22} to be truly one-dimensional,
$[u_y]\/$ (which incorporates the dissipation from the boundary layer)
must be expressed in terms of $\rho\/$ and $u^M\/$. As shown next in
\S~\ref{LAIC:ELM1:NoMo} and \ref{LAIC:ELM1:Re1D}, this can be done under
some restrictions.
%
%
\subsubsection{Normal modes}\label{LAIC:ELM1:NoMo}
Let $0 < \mu = \sqrt{\delta/\epsilon} \ll 1\/$. Then
(\ref{LAIC:ELM1:eqn:21}) takes the form \vspace*{-0.2em}
\begin{equation} \label{LAIC:ELM1:eqn:25}
 \rho_t + u^M_x = 0\/,\quad\mbox{and}\quad
 u_t + \rho_x = \mu^2\,u_{y\/y}\/,
 \quad\mbox{where}\quad u^M = \mbox{$\int$} u\,d\/y\/,
\end{equation}
where $u=0\/$ for $y = 0\/,\,1\/$. To show that \emph{these equations
include the results from the ``traditional'' approach,} we look for
solutions with exponential time dependence --- proportional to
$e^{-\ell^2\,t}\/$ ($\ell\/$ is a complex number). Then the second
equation in (\ref{LAIC:ELM1:eqn:25}) reduces to an ode BVP for $u\/$,
with solution: \vspace*{-0.2em}
\begin{equation} \label{LAIC:ELM1:eqn:26}
  u=-\tfrac{1}{\mu^2}\,G(\ell_*\/,\,y-\tfrac{1}{2})\,\rho_x\/,
\end{equation}
where\vspace*{-1.0em}
\begin{equation*}
  \ell_* = \frac{\ell}{\mu} \quad\mbox{and}\quad
  G(\ell_*\/,\,z) =
  \frac{\cos(\ell_*\,z)-\cos(\ell_*/2)}{\ell_*^2\,\cos(\ell_*/2)}\/.
\end{equation*}
The first equation in (\ref{LAIC:ELM1:eqn:25}) then reduces
to\,\footnote{\sf \,Note that $M\/$ is not singular for $\ell_*=0\/$,
   and that $M\/$ is a function of $\ell_*^2\/$.} \vspace*{-0.2em}
\begin{equation}\label{LAIC:ELM1:eqn:27}
 0 = M\,\rho_{x\/x} + \mu^2\ell^2\,\rho\/,
\end{equation}
where\vspace*{-1.0em}
\begin{equation*}
 M = \int_0^1 G(\ell_*\/,\,y-\tfrac{1}{2})\,d\/y
   = \frac{2}{\ell_*^3}\,\tan\left(\frac{\ell_*}{2}\right) -
     \frac{1}{\ell_*^2}\/.
\end{equation*}
Hence a longitudinal dependence proportional to $e^{i\,k\,x}\/$, $k\/$ a
real constant, yields the \emph{``dispersion'' relation}\vspace*{-0.2em}
\begin{equation}\label{LAIC:ELM1:eqn:28}
 k^2\,M(\ell_*) = \mu^2\,\ell^2\/.
\end{equation}
It is easy to see that, for $\ell_*\/$ with a large imaginary part,
$\ell_*^3\,M(\ell_*) \sim -\ell_*+2\,i\,\sigma\/$, where \emph{the error
is exponentially small} and $\sigma = \text{sign}(\text{Im}(\ell^*))\/$. 
Using this approximation in (\ref{LAIC:ELM1:eqn:28}), and solving for
$\ell\/$ in terms of $k\/$, yields \vspace*{-0.2em}
\begin{equation}\label{LAIC:ELM1:eqn:29}
 \ell = p_r\,\sqrt{|k|} -i\,\mu\,\tfrac{\sigma}{2} +
        {\mathcal O}\left(\tfrac{\mu^2}{\sqrt{|k|}}\right)\/,
\end{equation}
where $p_r = (\pm 1 +i\,\sigma)/\sqrt{2}\/$, $\sigma=\pm 1\/$,
is any one of the four roots of $p_r^4 = -1\/$. Hence \vspace*{-0.2em}
\begin{equation*}
 -\ell^2 = \pm\,i\,\sigma\,\left(|k|+\mu\,\sqrt{\tfrac{|k|}{2}}\right)
           -\mu\,\sqrt{\tfrac{|k|}{2}} + O(\mu^2)\/.\vspace*{-0.2em}
\end{equation*}
%
Since the time evolution is via $e^{-\ell^2\,t}\/$, equation
(\ref{LAIC:ELM1:eqn:29}) shows that the boundary layer corrections have
two effects:
\\(i)  a small frequency correction to the waves, given by
       $\mu\,\sqrt{|k|/2}\/$.
\\(ii) Dissipation, with dissipation coefficient
       $\mu\,\sqrt{|k|/2}\/$.

This approximation requires $|k| \gg \mu^2\/$. It is not valid for very
long waves, while (\ref{LAIC:ELM1:eqn:25}) remains valid beyond the wave
regime (very small $k\/$) --- see (\ref{LAIC:ELM1:eqn:23}). Note also
that (\ref{LAIC:ELM1:eqn:28}) has other modes/solutions in the regime
$|k|\gg\mu^2\/$, which are excluded by (\ref{LAIC:ELM1:eqn:29}). These
modes are not wave-like, and decay much faster than the ones retained.
\subsubsection{Reduction to a 1-D in space problem}
\label{LAIC:ELM1:Re1D}
Assume solutions such that (\ref{LAIC:ELM1:eqn:29}) applies --- i.e.,
enforce a very long wave frequency cut-off, and note that
(\ref{LAIC:ELM1:eqn:29}) is equivalent (up to the order displayed)
to \vspace*{-0.2em}
\begin{equation}\label{LAIC:ELM1:eqn:30}
 \left(-\ell^2 + \mu\,\sqrt{\tfrac{|k|}{2}}\right)^2 = -
 \left(|k| + \mu\,\sqrt{\tfrac{|k|}{2}}\right)^2\/.
\end{equation}
Introduce now the pseudo-differential operators (on functions of $x\/$)
defined by \vspace*{-0.2em}
\begin{eqnarray}
 {\mathcal L}_1\,e^{i\,k\,x} & = & i\,\text{sign}(k)\,
            \left(\mu\,\sqrt{2\,|k|}+\mu^2\right)\,e^{i\,k\,x}
            \nonumber \\ \mbox{and}\qquad
 {\mathcal L}_2\,e^{i\,k\,x} & = & -\mu\,\sqrt{2\,|k|}\,e^{i\,k\,x}\/.
 \label{LAIC:ELM1:eqn:31}
\end{eqnarray}
With these definitions
\begin{equation} \label{LAIC:ELM1:eqn:32}
 \rho_t + u^M_x = 0\/,\qquad\mbox{and}\qquad
 u^M_t + \rho_x = \mathcal L_2\,u^M - \mathcal L_1\,\rho,
\end{equation}
yields the dispersion relation (\ref{LAIC:ELM1:eqn:30}). Note that
(\ref{LAIC:ELM1:eqn:32}) is (\ref{LAIC:ELM1:eqn:22}) with
$\mu\,[u_y] = {\mathcal L}_2\,u^M - {\mathcal L}_1\,\rho\/$.
At this moment it is (to us) \emph{unclear if an equation like 
(\ref{LAIC:ELM1:eqn:32}) can be written that is fully equivalent to
(\ref{LAIC:ELM1:eqn:22})} --- no frequency cut-off.
Equations such as (\ref{LAIC:ELM1:eqn:32}) can only be formulated for
problems where the boundary conditions in $x\/$ are compatible with a
Fourier Expansion, such as full line or periodic. This limits their
usefulness.

A final note: Equations similar to (\ref{LAIC:ELM1:eqn:32}) can be found
in many papers and books dealing acoustic boundary layer
dissipation (e.g., see~\cite{Blackstock2000,Chester1964roct}). However,
quite often, the pseudo-differential operators employed operate in time,
not space. For example, convolution operators of the form
${\mathcal L}_*\,u = \int_0^\infty K(s)\,u(t-s)\,d\/s\/$, or defined via
how they operate on exponential functions of the form
$e^{i\,\omega\,t}\/$. There are a several \emph{fundamental problems
with such definitions:} (i) The situation under consideration involves
decay of the solutions in time, hence any possible extension of them
backwards in time will either blow up exponentially (at best) or not be
even possible (dissipation problems are often ill-posed for negative
time). Thus the operators so defined may have no meaning. (ii) Even if
a meaning exists, the problem is no longer a problem for which an IVP
makes sense, one needs to know the whole past to go forwards in time.
%

\section{Conclusion}
We presented a variation of matched asymptotic expansions where the
matching occurs at the equation level --- not their solutions. This
has the advantage that it does not require much information about the
inner and outer solutions in
order to perform the matching. Of course, the method also has the
obvious disadvantage that the equations obtained typically cannot be
solved analytically, and thus numerical tools must be employed.

%

\section*{Acknowledgments}
The research of R. R. Rosales was partially supported by NSF grants
DMS-1318942 and DMS-1614043.

\bibliographystyle{sapm}
\bibliography{lfaria-refs.bib}

\end{document}